\documentclass[12pt]{article}

\usepackage{amsmath,amssymb,epsfig}
\usepackage{graphicx}
\usepackage{arydshln}

\begin{document}
\setlength{\parskip}{0.45cm}
\setlength{\baselineskip}{0.75cm}

%
%
%
\begin{titlepage}
\setlength{\parskip}{0.25cm}
\setlength{\baselineskip}{0.25cm}
\begin{flushright}
DO-TH 2005/06\\
\vspace{0.2cm}
April 2005
\end{flushright}
\vspace{1.0cm}
\begin{center}
\Large
{\bf Signatures of cosmic tau--neutrinos}
\vspace{1.5cm}

\large
E.\ Reya and J.\ R\"odiger\\
\vspace{1.0cm}

\normalsize
{\it Universit\"{a}t Dortmund, Institut f\"{u}r Physik,}\\
{\it D-44221 Dortmund, Germany} \\
\vspace{0.5cm}

\vspace{1.5cm}
\end{center}

\begin{abstract}
\noindent The importance and signatures of cosmic tau--(anti)neutrinos
have been studied for upward-- and downward--going $\mu^-+\mu^+$ and 
hadronic shower event rates relevant for present and future underground
water or ice detectors, utilizing the unique and reliable ultrasmall--$x$
predictions of the dynamical (radiative) parton model.  
The upward--going $\mu^- +\mu^+$ event rates calculated just from cosmic
$\nu_{\mu}+\bar{\nu}_{\mu}$ fluxes are sizeably enhanced by taking into
account cosmic $\nu_{\tau}+ \bar{\nu}_{\tau}$ fluxes and their associated 
$\tau^- +\tau^+$ fluxes as well.  
The coupled transport
equations for the upward--going $\stackrel{(-)}{\nu}_{\tau}$ flux 
traversing the Earth imply an enhancement of the attenuated and regenerated
$\stackrel{(-)}{\nu}_{\tau}$ flux typically around $10^4-10^5$ GeV with
respect to the initial cosmic flux. This enhancement turns out to be smaller 
than 
obtained so far, in particular for flatter initial cosmic fluxes behaving
like $E_{\nu}^{-1}$.  Downward--going $\mu^- +\mu^+$ events and in particular
the background--free and unique hadronic  `double bang' and   `lollipop'
events allow to test downward--going cosmic $\nu_{\tau} +\bar{\nu}_{\tau}$
fluxes up to about $10^9$ GeV. 
\end{abstract}
\end{titlepage}


\section{Introduction}
The observation of cosmic high to ultrahigh energy neutrinos with energies
above 1 TeV is one of the important challenges of cosmic ray detectors in
order to probe the faintest regions of the Universe, i.e., astrophyics 
phenomena such as galaxy formation as well as particle (possibly `new')
physics.  The sources of cosmic (anti)neutrinos range, however, from the
well established to the highly speculative \cite{ref1,ref2,ref3,ref4,ref5},
such as active galactic nuclei (AGN) \cite{ref6,ref7,ref8}, gamma ray bursts
(GRB) \cite{ref9}, decays of exotic heavy particles of generic top--down
or topological defects (TD) \cite{ref10,ref11,ref12,ref13,ref14} and 
$Z$--bursts \cite{ref15,ref16,ref17,ref18}.  Representative fluxes of some 
of these hypothesized sources are displayed in Fig.~1 which we shall use,
as in \cite{ref19}, for all our subsequent calculations.  Although the 
rather prominent AGN--SS flux \cite{ref7} is in conflict with a recent
upper bound \cite{ref20} for $10^6$ GeV $\leq E_{\nu}<10^8$ GeV, we shall
keep using it for comparison with previous analyses. 

Apart from these violently different expectations for cosmic (anti)neutrino
fluxes, there are further uncertainties when calculating event rates for
neutrino telescopes.
A serious uncertainty is related to the sensitivity of 
$\stackrel{(-)}{\nu}\!\!N$
cross sections to the parton distributions at the weak scale $Q^2=M_W^2$
in the yet unmeasured Bjorken--$x$ region 
$x$ \raisebox{-0.1cm}{$\stackrel{<}{\sim}$} $10^{-3}$, in particular
their extrapolation to $x<10^{-5}$ as soon as 
$E_{\nu}$ \raisebox{-0.1cm}{$\stackrel{>}{\sim}$} $10^8$ GeV
in Fig.~1 ($x\simeq M_W^2/2M_NE_{\nu})$.  Leaving aside somewhat arbitrary
extrapolation techniques based on assumptions on various fixed power
behaviors in $x$ of structure functions as $x\to 0$ 
\cite{ref21,ref22,ref23,ref24}, such extensive small--$x$ extrapolations
can be performed more reliably by using the QCD inspired dynamical 
(radiative) parton model \cite{ref25} which proved to provide reliable
deep inelastic high energy predictions in the past (a more detailed discussion
and summary can be found in \cite{ref19}).  Within this approach the entire
partonic structure at 
$x$ \raisebox{-0.1cm}{$\stackrel{<}{\sim}$} $10^{-2}$ can be understood
and calculated via renormalization group evolutions from first principles,
i.e., QCD dynamics, independently of free (fit) parameters in the small--$x$
region.  It has furthermore been shown \cite{ref26} that 
(anti)neutrino--nucleon
cross sections can be calculated with an uncertainty of about $\pm 20\%$
at highest neutrino energies of $10^{12}$ GeV. (The relevant
cross sections obtained from the fitted CTEQ3--DIS parametrizations
\cite{ref27} at 
$x$ \raisebox{-0.1cm}{$\stackrel{>}{\sim}$} $10^{-5}$ with their
assumed fixed--power extrapolation to $x<10^{-5}$ accidentally coincide
practically with the ones derived from the dynamical ultrasmall--$x$
predictions of the radiative parton model \cite{ref25}; these `variable
flavor'  CTEQ3--DIS densities, where the heavy $c,\, b,\, t$ quarks are
effectively treated as massless intrinsic partons, are easier to use for
practical calculations).  These dynamical small--$x$ predictions have been
recently utilized for recalculating \cite{ref19} muon event rates produced
by (mainly) upward--going muon--neutrinos with energies below $10^8$ GeV
\cite{ref21,ref22} in large--volume underground water or ice detectors
(AMANDA/IceCube, ANTARES, NESTOR, NEMO \cite{ref2,ref28}).  When 
penetrating through the Earth, the cosmic muon (anti)neutrinos undergo 
attenuation (absorption) due to charged current (CC) and neutral current 
(NC) interactions as well as regeneration \cite{ref29,ref30} due to the
NC interactions which shift their energy, rather than absorbing them, to
lower energies and populate the lower energy part of the initial flux
spectra shown in Fig.~1 \cite{ref19,ref31}, thus increasing the naive
non--regenerated $\mu^- +\mu^+$ event rates.

It is our main objective to extend and complete the previous analysis 
\cite{ref19} by taking into account cosmic tau--(anti)neutrinos along the
line of \cite{ref32,ref33}.  Due to near--maximal
$\nu_{\mu}-\nu_{\tau}$~mixing \cite{ref34,ref35,ref36}, the 
$\nu_{\tau}+\bar{\nu}_{\tau}$ flux arriving at the Earth's surface equals
the $\nu_{\mu}+\bar{\nu}_{\mu}$ flux \cite{ref37} 
and thus may significantly enhance
the upward $\mu^-+\mu^+$ and (hadronic) shower event rates according
to their interaction in Earth \cite{ref32,ref33,ref38,ref39,ref40} via
$\nu_{\tau}N\to\tau X\to\mu X'$, etc.  Because of the latter (semi)leptonic
decay $\tau\to\nu_{\tau} X$, the Earth never becomes opaque to tau--neutrinos
as long as the interaction lengths of the taus is larger than their decay
length (which holds for energies up to about $10^9$ GeV),
in contrast to muon-- and electron--neutrinos \cite{ref41}:  a high--energy
$\nu_{\tau}$ interacts in the Earth producing taus which, due to the short
lifetime, in turn decay into a $\nu_{\tau}$ with lower energy.  This
`regeneration chain' $\nu_{\tau}\to\tau\to\nu_{\tau}\to\ldots$ continues
until the $\tau$--neutrinos (as well as the $\tau$--leptons) reach the detector
on the opposite side of the Earth.  Thus the propagation of high--energy
tau--neutrinos through the Earth is very different from muon-- and 
electron--neutrinos.  Instead of a single transport (integro--differential)
equation for muon--neutrinos \cite{ref29,ref30,ref31,ref19} we have now to
deal with coupled transport equations for the $\nu_{\tau}$ and $\tau$ fluxes.
This will be done in Sect.~2 and the resulting fluxes presented.  Since we
do not fully confirm the results obtained for the $\nu_{\tau}+\bar{\nu}_{\tau}$
flux in the literature, a detailed derivation of the solutions of the most
general transport equations is given in the Appendix, together with the
resulting approximations relevant for our calculations in order to keep this
paper as far as possible self--contained.  The appropriate upward
$\mu^-+\mu^+$ event rates, being the most numerous in modern underground
detectors \cite{ref33,ref42}, are presented in Sect.~3.  In particular,
the additional and sizeable contributions arising from the $\tau^-+\tau^+$
flux, generated by the initial cosmic $\nu_{\tau}+\bar{\nu}_{\tau}$ flux
when traversing the Earth, will be calculated as well which so far have been
disregarded when calculating upward--going $\mu^-+\mu^+$ event rates.

For neutrino energies above $10^5$ GeV the shadowing in Earth rapidly 
increases which severely restricts rates in underground detectors 
\cite{ref19,ref22,ref24,ref33,ref42}.  Eventually it becomes beneficial to 
look for events induced by downward--going and (quasi)horizontal neutrinos 
\cite{ref19,ref22,ref24,ref33,ref42,ref43,ref44}, provided of course such 
events produced by interactions within the instrumented underground detector 
volume can be efficiently observed.  So--called `double bang' and `lollipop' 
events \cite{ref35} are signatures unique to tau--neutrinos which seem to 
be most promising to recognize $\tau$--leptons \cite{ref45,ref42,ref40}. 
A  double bang event consists of a hadronic shower initiated by the
$\stackrel{(-)}{\nu}\!\!\!_{\tau}N$ CC interaction vertex followed by a second
energetic hadronic (or electromagnetic) shower due to the decaying tau.  
A lollipop event consists
only of the second of the two showers along with the 
reconstructed $\tau$--lepton track and with the first shower at the 
CC interaction vertex outside of the sensitive detector volume.  The
relevant downward rates, being far more sensitive to the specific choice 
of parton distributions than the upward--going rates \cite{ref19,ref21}, 
will be presented in Sect.~4 and compared with the ones resulting from 
downward--going muon--neutrinos calculated previously \cite{ref19}. It
should be emphasized that tau--neutrinos offer an ideal means of 
identifying neutrinos of {\underline{cosmic}} origin (and for searching
for possible `new' physics) since the conventional atmospheric flux
background is negligible for $E_{\nu}>10^3$ GeV \cite{ref46,ref47}, in
contrast to muon--neutrinos \cite{ref48,ref46}; furthermore the flux of prompt
$\nu_{\tau}$ neutrinos (from charm and bottom production, hadronization
and decay) is about ten times less than for prompt $\nu_{\mu}$ neutrinos
\cite{ref46,ref47}.  

At energies above $10^8$ GeV where the (anti)neutrino interaction length
becomes smaller than $10^3$ km water equivalent (we) in rock, upward--going
neutrinos are blocked by Earth and thus underground detectors become
ineffective due to the opaqueness of Earth to upward--going neutrinos.
Therefore large--area ground arrays or surface fluorescence telescopes
such as AGASA, the HiRes detector (an upgrade of Fly's Eye) and the 
Pierre Auger Observatory \cite{ref43,ref49,ref50}, or antarctic balloon
missions (ANITA) \cite{ref51} will be instrumental in exploring the spectrum
of cosmic neutrino fluxes up to highest energies of about $10^{12}$ GeV
shown in Fig.~1.  Here the interaction medium, which acts as neutrino
converter, is either the atmosphere or more effectively the Earth's crust
or ice.  In particular tau--neutrinos $\nu_{\tau}+\bar{\nu}_{\tau}$ when
skimming the Earth \cite{ref52,ref53}, i.e.\ entering Earth near--horizontally 
at some large critical nadir angle 
$\theta$ \raisebox{-0.1cm}{$\stackrel{>}{\sim}$} $85^{\rm o}$, 
are most effective
in producing lollipop and double bang events, including electromagnetic
showers.  These effects and resulting rates have been extensively studied
in the past \cite{ref19,ref39,ref44,ref53,ref54,ref55,ref56,ref57,ref58}
and will not be considered any further.

\section{Propagation of $\stackrel{(-)}{\nu}\!\!\!_{\tau}$ and $\tau^{\pm}$
through the Earth}
The transport equation for muon--neutrinos \cite{ref29,ref30} is
straightforwardly generalized to the coupled transport equations relevant
for tau--(anti)neutrinos and tau--leptons:  for $\stackrel{(-)}{\nu}\!\!\!_{\tau}$
one has to take into account the attenuation due to 
$\sigma_{\nu_{\tau}N}^{\rm tot}=\sigma_{\nu_{\tau}N}^{\rm CC}
+ \sigma_{\nu_{\tau}N}^{\rm NC}$, and the regeneration consisting of
the degrading shift in neutrino energy due to $\sigma_{\nu_{\tau}N}^{\rm NC}$
and of $\sigma_{\tau N}^{\rm CC}$ as well as of the $\tau$--decay
when $\stackrel{(-)}{\nu}\!\!\!_{\tau}$ and $\tau^{\pm}$ penetrate through the
Earth. The latter tau--leptons are produced in CC interactions via
$\sigma_{\nu_{\tau}N}^{\rm CC}$ and attenuated via their decay and
CC interaction $\sigma_{\tau{N}}^{\rm CC}$; in general one also has to
include the electromagnetic energy loss of $\tau^{\pm}$ as well. The 
resulting coupled transport equations for the fluxes of tau--(anti)neutrinos
and tau--leptons are given by
\renewcommand{\arraystretch}{3.0}
\begin{eqnarray}
\frac{\partial F_{\nu_{\tau}}(E,X)}{\partial X} 
 & = & -\frac{F_{\nu_{\tau}}(E,X)}{\lambda_{\nu_{\tau}}(E)} 
    + \frac{1}{\lambda_{\nu_{\tau}}(E)} \int_0^1\frac{dy}{1-y}\,
      K_{\nu_{\tau}}^{\rm NC}(E,y)\, F_{\nu_{\tau}}(E_y,X) \nonumber\\
& & + \int_0^1\frac{dy}{1-y}\, K_{\tau}(E,y)F_{\tau}(E_y,X)\\
\nonumber\\
\frac{\partial F_{\tau}(E,X)}{\partial X} 
 &  = & - \frac{F_{\tau}(E,X)}{\hat{\lambda}_{\tau}(E)} + 
       \frac{\partial\left[\gamma(E)F_{\tau}(E,X)\right]}{\partial E}\nonumber\\
 & &  + \frac{1}{\lambda_{\nu_{\tau}}(E)} \int_0^1 \frac{dy}{1-y}\,
      K_{\nu_{\tau}}^{\rm CC} (E,y)\, F_{\nu_{\tau}}(E_y,X)
\end{eqnarray}

\noindent
where $F_{\nu_{\tau}}=d\Phi_{\nu_{\tau}}/dE$ and $F_{\tau}=d\Phi_{\tau}/dE$
are the differential energy spectra (fluxes) of (anti)tau--neutrinos 
(cf.~Fig.~1) and $\tau^{\pm}$ leptons.  The column depth $X=X(\theta)$,
being the thickness of matter traversed by the upgoing leptons, depends
upon the nadir angle of the incident neutrino beam ($\theta=0^{\rm o}$
corresponds to a beam traversing the diameter of the Earth); it is obtained
from integrating the density $\rho(r)$ of the Earth along the neutrino
beam path $L'$ at a given $\theta$, $X(\theta) = \int_0^L\rho(L')dL'$ with
$L=2R_{\oplus}\cos\theta$ denoting the position of the underground detector,
and $X(\theta)$ is given in Fig.~15 of \cite{ref21} in units of 
g/cm$^2=$ cm~we.  Furthermore         
$\lambda_{\nu_{\tau}}^{-1} = N_A \, \sigma_{\nu_{\tau}N}^{\rm tot}$,       
$\hat{\lambda}_{\tau}^{-1} = (\lambda_{\tau}^{\rm CC})^{-1} +
(\lambda_{\tau}^{\rm dec})^{-1}$ with $(\lambda_{\tau}^{\rm CC})^{-1} =
N_A\, \sigma_{\tau N}^{\rm CC}$ and $N_A = 6.022\times 10^{23}\,{\rm g}^{-1}$, and
the decay length of the $\tau^{\pm}$ is 
$\lambda_{\tau}^{\rm dec}(E,X,\theta) = (E/m_{\tau})c\tau_{\tau}\rho$
with $m_{\tau}=1777$ MeV, $c\tau_{\tau}=87.11$ $\mu$m and in order to
simplify \cite{ref32} the solution of (2) for $F_{\tau}$ one uses the
reasonable approximation $\rho(X,\theta)\simeq \rho_{av}(\theta)$ where
the average of the Earth's density along the column depth is calculated
according to $\rho_{av}(\theta)=X(\theta)/L=X(\theta)/2R_{\oplus}\cos\theta$,
with $R_{\oplus}\simeq 6371$ km. Thus 
$\lambda_{\tau}^{\rm dec}(E,X,\theta)\simeq \lambda_{\tau}^{\rm dec}(E, \theta)
= (E/m_{\tau})c\tau_{\tau}\rho_{av}(\theta)$.
Note that a possible contribution from $\sigma_{\tau N}^{\rm NC}$ has been
disregarded in (2) since the second term on the r.h.s.\ of (2), describing
the electromagnetic energy--loss of $\tau^{\pm}$ leptons proportional to 
$\gamma(E)=\alpha_{\tau}(E)+\beta_{\tau}(E)E$, dominates for 
$E$ \raisebox{-0.1cm}{$\stackrel{<}{\sim}$} $10^{16}$ GeV \cite{ref44,ref59}.
The remaining cross section and decay kernels in (1) and (2) are given by
\renewcommand{\arraystretch}{3.0}
\begin{eqnarray}
K_{\nu_{\tau}}^{\rm NC,\, CC}(E,y) & = & \frac{1}{\sigma_{\nu_{\tau}N}^{\rm tot}(E)}\,
   \frac{d\sigma_{\nu_{\tau}N}^{\rm NC,\, CC}(E_y,y)}{dy}\,\, ,\nonumber\\
K_{\tau}^{\rm CC}(E,y) & = & \frac{1}{\sigma_{\tau N}^{\rm tot}(E)}\, 
   \frac{d\sigma_{\tau N}^{\rm CC}(E_y,y)}{dy}\, \, , K_{\tau}^{\rm dec}(E,y) 
    =\frac{1}{\Gamma_{\tau}^{\rm tot}(E)} \, 
       \frac{d\Gamma_{\tau\to \nu_{\tau}X}(E_y,y)}{dy}\nonumber\\
K_{\tau}(E,y)& = & \frac{1}{\lambda_{\tau}(E)}\, K_{\tau}^{\rm CC}(E,y) +
   \frac{1}{\lambda_{\tau}^{\rm dec}(E)}\, K_{\tau}^{\rm dec}(E,y)
\end{eqnarray}

\noindent
where $E_y = E/(1-y)$, $\lambda_{\tau}^{-1}=N_A\, \sigma_{\tau N}^{\rm tot}$ and the
obvious dependence on the nadir angle $\theta$, like in $\lambda^{\rm dec}_{\tau}$,
will be suppressed from now on.  The various CC and NC 
$\stackrel{(-)}{\nu}\!\!\!_{\tau}N$ cross sections are calculated as in \cite{ref19},
with the details to be found in \cite{ref26}, utilizing the dynamical small--$x$
predictions for parton distributions according to the radiative parton model
\cite{ref25}.  Furthermore, since $1/\Gamma_{\tau}^{\rm tot}(E)=(E/m_{\tau})
\tau_{\tau}$, we have more explicitly for the $\tau$--decay distribution 
$K_{\tau}^{\rm dec}(E,y)=(1-y)dn(z)/dy$ with $z=E_{\nu_{\tau}}/E_{\tau} =
E/E_y = 1-y$ and \cite{ref33,ref60}
\begin{equation}
\frac{dn(z)}{dy} =\sum_i B_i \left[g_0^i(z)+P\, g_1^i(z)\right]
\end{equation}
with the polarization $P=\pm 1$ of the decaying
$\tau^{\pm}$ and where the $\tau\to\nu_{\tau}X$ branching fractions $B_i$ 
into
the decay channel $i$ and $g_{0,1}^i(z)$ are given in Table I of \cite{ref33}.
An equation similar to (2) has been found in \cite{ref61} in the context of
atmospheric muons where the lepton energy--loss is treated continuously, i.e.
by the term proportional to $\gamma(E)$.  In contrast to muons, this continuous
approach of the energy--loss of taus does not significantly overestimate the 
tau--range \cite{ref44} as compared to treating the average energy--loss separately
(stochastically) \cite{ref57,ref59}, i.e. not including the term proportional
to $\gamma(E)$ in (2) but using instead 
$-dE_{\tau}/dX=\gamma(E_{\tau})=\alpha_{\tau}+\beta_{\tau}E_{\tau}$.
For definiteness all above formulae have been given for an incoming neutrino
beam, but similar expressions hold of course for antineutrinos.

The general (iterative) solution of the coupled transport equations (1) and
(2) will be, for completeness, derived in the Appendix.  For our purpose,
however, it suffices to work with the following simplifying assumptions for
energies smaller than $10^8$ GeV relevant for upward--going neutrinos: here
the $\tau^{\pm}$ energy--loss $\gamma(E_{\tau})$ can be neglected 
\cite{ref44,ref57,ref59,ref38} and the tau--lepton interaction length is (much)
larger than the decay length of the $\tau$ \cite{ref44,ref57,ref59}. In 
other words, for $E<10^8$ GeV, 
\begin{equation}
\gamma(E)\simeq 0,\quad\quad \lambda_{\tau}(E)\gg \lambda_{\tau}^{\rm dec}(E)
\end{equation}
i.e. $K_{\tau}(E,y)\simeq K_{\tau}^{\rm dec}(E,y)/\lambda_{\tau}^{\rm dec}(E)$
in (1) and, besides neglecting the term $\partial[\gamma F_{\tau}]/\partial E$ 
in (2),
$\hat{\lambda}_{\tau}^{-1}\simeq(\lambda_{\tau}^{\rm dec})^{-1}$.  With these
approximations, the solutions of Eqs.~(1) and (2), after a sufficiently accurate
first iteration (see Appendix), become
\begin{equation}
F_{\nu_{\tau}}(E,X) = F_{\nu_{\tau}}^0(E)\exp\left\{ -\frac{X}
  {\Lambda_{\nu}^{(1)}(E,X)}\right\}
\end{equation}
\begin{equation}
F_{\tau}(E,X)=\frac{F_{\nu_{\tau}}^0(E)}{\lambda_{\nu_{\tau}}(E)}\,
  e^{-\frac{X}{\lambda_{\tau}^{\rm dec}(E)}}
   \int_0^X\!\! dX'\int_0^1\!\! dy\, K_{\nu_{\tau}}^{\rm CC}
                                          (E,y)\eta_{\nu_{\tau}}(E,y)\,\,
     e^{-\frac{X'}{\Lambda_{\nu}^{(1)}(E_y,X')}}\,\,
       e^{\frac{X'}{\lambda_{\tau}^{\rm dec}(E)}}
\end{equation}

\noindent
with $\Lambda_{\nu}^{(1)}(E,X)=\lambda_{\nu_{\tau}}(E)/[1-Z^{(1)}(E,X)]$
where $Z^{(1)}=Z_\nu^{(1)}+Z_{\tau}^{(1)}$ with
\vspace{-0.75cm}

\begin{eqnarray}
Z_{\nu}^{(1)}(E,X) & = & \int_0^1 dy\, K_{\nu_{\tau}}^{\rm NC}(E,y)
  \eta_{\nu_{\tau}}(E,y)\, \frac{1-e^{-X D_{\nu}(E,E_y)}}{X D_{\nu}(E,E_y)}
 \nonumber\\
Z_{\tau}^{(1)}(E,X) & = & \frac{\lambda_{\nu_{\tau}}(E)}
   {\lambda_{\tau}^{\rm dec}(E)} \int_0^1 dy\int_0^1 dy'\, 
    K_{\tau}^{\rm dec}(E,y)\, K_{\nu_{\tau}}^{\rm CC}(E_y,y')
\lambda_{\nu_{\tau}}^{-1}(E_y)\eta_{\nu_{\tau}}(E,y)\eta_{\nu_{\tau}}(E_y,y')
\nonumber\\
& & \times \frac{1}{XD_{\nu\tau}(E_y,E_{yy'})}\, 
    \Bigg\{\frac{1}{D_{\tau\nu}(E,E_y)}
      \left(1-e^{-XD_{\tau\nu}(E,E_y)}\right)\nonumber\\
& &  -\frac{1}{D_{\nu}(E,E_{yy'})}
         \left(1-e^{-X D_{\nu}(E,E_{yy'})}\right) \Bigg\} 
\end{eqnarray}

\noindent
where $E_{yy'} = E_y/(1-y')=E/(1-y)(1-y')$ and
\begin{equation}
D_{\nu}(E,E_y) = \frac{1}{\lambda_{\nu_{\tau}}(E_y)} -
   \frac{1}{\lambda_{\nu_{\tau}}(E)}\, ,
  D_{\nu\tau}(E,E_y) = \frac{1}{\lambda_{\nu_{\tau}}(E_y)} - 
   \frac{1}{\lambda_{\tau}^{\rm dec}(E)}\, , 
  D_{\tau\nu}(E,E_y) = - D_{\nu\tau}(E_y,E)\, .
\end{equation}

\noindent
Furthermore, $\eta_{\nu_{\tau}}(E,y) = F^0_{\nu_{\tau}}(E_y)/(1-y)
F_{\nu_{\tau}}^0(E)$ with the initial cosmic neutrino flux which reaches 
the Earth's surface being denoted by $F_{\nu_{\tau}}^0(E)=
F_{\nu_{\tau}}(E,\, X=0)$. Note that $F_{\nu_{\tau}}^0(E) = 
F_{\bar{\nu}_{\tau}}^0(E) = \frac{1}{4}d\Phi/dE$ with $\Phi$ being the cosmic 
$\nu_{\mu}+\bar{\nu}_{\mu}$ flux in Fig.~1.

For a better comparison of our quantitative upward--going flux results
with the ones obtained in the literature, we employ two generic initial 
fluxes of the form \cite{ref32, ref33}
\begin{eqnarray}
F_{\nu_{\tau}+\bar{\nu}_{\tau}}^0(E_{\nu}) & = & N_1 E_{\nu}^{-1}
   (1+E_{\nu}/E_0)^{-2},\quad E_0 = 10^8\,{\rm GeV}\\
F_{\nu_{\tau} +\bar{\nu}_{\tau}}^0(E_{\nu}) & = & N_2 E_{\nu}^{-2}
\end{eqnarray}
with adjustable normalization factors $N_i$, for example, 
$N_1=\frac{1}{2}\times 10^{-13}$/(cm$^2$ sr s) and 
$N_2=\frac{1}{2}\times 10^{-7}$
GeV/(cm$^2$ sr s).  Notice that the generic $E_{\nu}^{-1}$ energy dependence 
is representative for the TD and $Z$--burst fluxes in Fig.~1 for 
$E_{\nu}$ \raisebox{-0.1cm}{$\stackrel{<}{\sim}$} $ 10^7$ GeV, 
and partly also for the AGN-SS flux, as well 
as for the GRB--WB flux for 
$E_{\nu}$ \raisebox{-0.1cm}{$\stackrel{<}{\sim}$} $10^5$ GeV; 
furthermore the latter GRB--WB flux behaves like $E_{\nu}^{-2}$ in (11) for 
$10^5<E_{\nu}$ \raisebox{-0.1cm}{$\stackrel{<}{\sim}$} $10^7$ GeV.
Our results are shown in Fig.~2 and compared with the ones of \cite{ref33}.
The typical enhancement (`bump') of the attenuated and regenerated
$\stackrel{(-)}{\nu}\!\!\!_{\tau}$ flux around $10^4-10^5$ GeV, which is 
prominent for the flatter 
$F_{\stackrel{(-)}{\nu}\!\!_{\tau}}^0\sim E_{\nu}^{-1}$
flux and absent for a $\stackrel{(-)}{\nu}\!\!\!_{\mu}$ flux, amounts to about
40\% with respect to the initial neutrino flux (dashed curve) whereas the
results of \cite{ref33} amount to an enhancement of about a factor of 2.
It should be emphasized that our results are practically insensitive to
the high energy cutoff $E_0$ in (10).  This is in contrast to a Monte
Carlo simulation \cite{ref38} where an enhancement of a factor of 4 has
been found with respect to the initial $E_{\nu}^{-1}$ flux; however, it 
has been stated that it reduces to the result of \cite{ref33} if the high
energy cut--off in (10) is taken into account.  Such an enhanced bump
disappears for steeper fluxes like in (11) and the even steeper AGN--M95
flux in Fig.~1.  Here our results differ by less than 10\% from the ones
of \cite{ref33} as shown in Fig.~2.  The ratios of our results and the 
ones in \cite{ref32,ref33,ref62} are, for better illustration, plotted
in Fig.~3.  Since our results deviate rather sizeably from the ones in
\cite{ref32,ref33,ref62} for the flatter initial cosmic tau--neutrino fluxes
behaving like $E_{\nu}^{-1}$, the corresponding rates for upward--going
$\mu^{\pm}$ and shower events will be, on the average, about half as large
than in [33].

The enhancement due to regeneration, typical for tau--(anti)neutrinos,
relative to the initial $\nu_{\tau}+\bar{\nu}_{\tau}$ fluxes in Fig.~1
is illustrated in Fig.~4 for $\theta =0^{\rm o}$ and 30$^{\rm o}$ (remember
that $\theta = 0^{\rm o}$ corresponds to a beam traversing the diameter of the
Earth).  
This effect is prominent for flatter initial fluxes $\sim E_{\nu}^{-1}$ 
whereas it is absent for steeper fluxes $\sim E_{\nu}^{-n}$, 
\mbox{$n$ \raisebox{-0.1cm}{$\stackrel{>}{\sim}$} 2}, 
like the AGN-M95 flux for which the ratios in Fig.~4 are always smaller 
than 1.  It is equally absent for $\stackrel{(-)}{\nu}\!\!\!_{\mu}$ fluxes 
\cite{ref19,ref31,ref32,ref33} where no decay contribution exists in the 
transport equation. Finally, the results for the absolute 
$\nu_{\tau}+\bar{\nu}_{\tau}$ fluxes and the $\tau^- +\tau^+$ fluxes, 
arising from the initial $\nu_{\tau}+\bar{\nu}_{\tau}$ fluxes, are presented 
in Fig.~5.  The $\nu_{\tau} + \bar{\nu}_{\tau}$ results correspond of 
course to the relative ratios shown in Fig.~4.  The $\tau^- + \tau^+$
fluxes at the detector site, despite being (superficially) suppressed
with respect to the $\nu_{\tau} +\bar{\nu}_{\tau}$ fluxes, will sizeably
contribute to the upward--going $\mu^- +\mu^+$ and shower event rates.

\section{Upward muon event rates}
The upward--muon ($\mu^{\pm}$) event rate produced by an upward--going
$\stackrel{(-)}{\nu}\!\!\!_{\tau}$ can be easily obtained by modifying the 
standard formula for the muon rate produced by the upward--going
$\stackrel{(-)}{\nu}\!\!\!_{\mu}$ \cite{ref21}, by taking into account 
the decay of the $\tau$ produced by the CC interaction $\nu_{\tau}N\to
\tau X$ \cite{ref33}.  This decay distribution and branching fraction
for the $\tau\to\nu_{\tau}\nu_{\mu}\, \mu$ decay is given by (4) according
to 
$dn_{\tau^{\pm}\to\mu^{\pm}X}(z)/dz = B_{\mu}[g_0^{\mu}(z)\pm g_1^{\mu}(z)]$ 
where $B_{\mu}=0.18$ and $z=E_{\mu}/E_{\tau}$.   
Thus the $\stackrel{(-)}{\nu}\!\!\!_{\tau}$ initiated 
$\mu^{\stackrel{(+)}{-}}$ event rate
per unit solid angle and second is given by    
\begin{eqnarray}
N_{\mu^-}^{(\nu_{\tau})} & = & N_A\int_{E_{\mu}^{\rm min}}dE_{\nu}
  \int_0^{1-E_{\mu}^{\rm min}/E_{\nu}}dy
   \int_{E_{\mu}^{\rm min}/(1-y)E_{\nu}}^1
         dz\, A(E_\mu)\, R
    \left((1-y)zE_{\nu},\, E_\mu^{\rm min}\right)\nonumber\\
& &   \times \frac{dn_{\tau^-\to\mu^- X}(z)}{dz}\,\,
          \frac{d\sigma_{\nu_{\tau}N}^{\rm CC}(E_{\nu} ,y)}{dy}
        F_{\nu_\tau}(E_\nu ,X)
\end{eqnarray}

\noindent
where the energy dependent area $A(E_{\mu})$, $E_{\mu}=(1-y)zE_{\nu}$,
of the underground detector is taken as summarized in \cite{ref19}.
The range $R(E_{\mu},E_{\mu}^{\rm min})$ of an energetic $\mu^{\pm}$
being produced with energy $E_{\mu}$ and, as it passes through the 
medium (Earth) loses energy, arrives in the detector with an energy
above $E_\mu^{\rm min}$, follows from the energy--loss relation 
$-d E_\mu/d X = \alpha_\mu+\beta_\mu E_\mu$, i.e.,
\begin{equation}
R(E_\mu,\, E_\mu^{\rm min}) \equiv X(E_\mu^{\rm min})-X(E_\mu)
  = \frac{1}{\beta_\mu}\ln 
   \frac{\alpha_\mu+\beta_\mu E_\mu}{\alpha_\mu+\beta_\mu E_\mu^{\rm min}}
\end{equation}
with $\alpha_\mu = 2\times 10^{-3}$ GeV (cm we)$^{-1}$ and $\beta_\mu =
6 \times 10^{-6}$ (cm we)$^{-1}$ which reproduce very well \cite{ref19}
the Monte Carlo range result of Lipari and Stanev \cite{ref63}.  
Similarly, the upward--$\mu$ event rate per unit solid angle and second
produced by the upward--going $\tau$--flux in (7) becomes
\begin{equation}
N_\mu^{(\tau)} = \int_{E_\mu^{\rm min}} dE_\tau 
   \int_{E_\mu^{\rm min}/E_\tau}^1 dz\, A(E_\mu)\, 
      R(zE_\tau , E_\mu^{\rm min})
       \frac{1}{\lambda_\tau^{\rm dec}(E_\tau)}\, 
        \frac{dn_{\tau\to\mu X}(z)}{dz}\, F_\tau(E_\tau,X)
\end{equation}

\noindent
where $E_\mu = zE_\tau$.  Apart from Monte Carlo studies of the rates of
$\tau^{\pm}$ emerging from the Earth's surface \cite{ref38}, the contributions to
the $\mu^{\pm}$ event rates arising from the upward--going $\tau^{\pm}$
flux $F_\tau$ have not been taken into account so far.  For our practical
purposes the upper limits for the energy--integrations in (12) and (14)
will be taken to be $10^8$ GeV.  Furthermore in order to obtain the
important {\underline{total}} nadir angle integrated upward event rates,
(12) and (14) have to be integrated over $\int_0^{2\pi}d\varphi
\int_0^{\pi/2} d\theta\sin\theta = 2\pi\int_0^{\pi/2}d\theta\sin\theta$.

For completeness it should be mentioned that the fluxes of secondary 
$\bar{\nu}_e$ and $\bar{\nu}_\mu$, created by the prompt leptonic tau decays
$\tau\to\nu_\tau\, e\bar{\nu}_e$ and 
$\tau\to \nu_\tau\, \mu\bar{\nu}_{\mu}$,
may enhance the detectability of the initial cosmic $\nu_\tau$ flux
\cite{ref64}.  It has been shown, however, that the associated total
$\mu^- +\mu^+$ event rate will be difficult to observe experimentally 
\cite{ref62}.
Furthermore, the hadronic decay channels of the tau--lepton may also
enhance the $\mu^- +\mu^+$ event rates in (12) and (14).  The only
conceivable potentially competing hadronic decay channel would be 
$\tau\to \nu_{\tau}\pi$.  However, its branching fraction is only about
half as large as the purely leptonic one in (12)
and (14) and, moreover, the $\mu$ in the cascade decay $\pi\to\mu\nu_{\mu}$
will be degraded in energy.  Therefore such suppressed contributions 
have not been taken into account.

The total event rates as a function of $E_\mu^{\rm min}$ are shown in
Fig.~6 by the solid curves for the initial cosmic fluxes
in Fig.~1, whereas the corresponding rates in (12) arising just from the
$\nu_\tau+\bar{\nu}_\tau$ flux arriving at the detector, 
$F_{\nu_\tau +\bar{\nu}_\tau}(E_\nu ,X)$, are shown by the dashed curves.
We refrain from showing the upward--going event rates caused by the 
TD--SLSC and $Z$--burst fluxes in Fig.~1, since they are too small for any
realistic purpose.  In any case the upward--going $\tau$--flux
$F_{\tau^- +\tau^+}(E_\tau ,X)$ in (14) almost doubles the rates
initiated by the $\nu_\tau +\bar{\nu}_\tau$ flux.  
This is not entirely surprising despite the fact that the $\tau^- +\tau^+$
fluxes in Fig.~5 are up to about 10 orders of magnitude smaller than the
$\nu_\tau +\bar{\nu}_\tau$ fluxes, since the latter ones have to undergo
CC interactions for giving rise to the observable muons (cf.~(12)) in 
contrast to the taus in (14).
Adding these results
to the ones arising from the $\nu_\mu +\bar{\nu}_\mu$ fluxes
\cite{ref19} one obtains the total annual event rates as given in 
Table 1 where the latter $\nu_\mu +\bar{\nu}_\mu$ initiated rates 
\cite{ref19} are displayed in parentheses.  The additional $\mu^- +
\mu^+$ events arising from the  
$\nu_\tau+\bar{\nu}_\tau$ and $\tau^- +\tau^+$ fluxes in (12) and (14),
respectively, increase the $\nu_\mu +\bar{\nu}_\mu$ induced event rates
by typically 30 to 20\% for $E_{\mu}^{\rm min} = 10^3$ to $10^4$ GeV
and the enhancement is less pronounced (20 to 10\%) at higher energies.
This different energy (and nadir angle) dependence of upward--going
$\mu^- +\mu^+$ events, as well as of hadronic and electromagnetic shower
events, may signal the appearance of a cosmic $\nu_\tau +\bar{\nu}_\tau$
flux \cite{ref33,ref42} and its associated $\tau^- +\tau^+$ flux. 
Different energy--loss properties of $\mu$-- and $\tau$--leptons may
also serve as an indirect signature of the 
$\stackrel{(-)}{\nu}\!\!\!_\tau$ appearance \cite{ref40}.
Notice that (possibly energy dependent) detector efficiencies have not
been included in our calculations which are an intrinsic experimental
matter.  In case future measurements will require such corrections, the
rates could be easily recalculated once realistic efficiencies are 
provided by the experimentalists.

The contribution to the total event rates in Table 1 from energies
above $10^8$ GeV becomes, however, negligible and unmeasurably small due
to the reduction of the initial $\nu_\mu +\bar{\nu}_\mu$ and $\nu_\tau
+\bar{\nu}_\tau$ fluxes, and the associated $\tau^- +\tau^+$ flux, by
attenuation with or without regeneration \cite{ref19}, cf.~Fig.~5.
The highest event rates arise in the AGN models which might be testable
for neutrino flux energies as large as $10^7 - 10^8$ GeV, i.e. 
$E_\mu^{\rm min}=10^7$ GeV.  It should be kept in mind, however, that
the AGN-SS flux is already disfavored by experiment \cite{ref20}.   
Beyond neutrino energies of $10^8$ GeV present models of cosmic neutrino
fluxes are not testable anymore by upward--going $\mu^- +\mu^+$ events.
Notice that the atmospheric (ATM) neutrino background, due to the dominant
$\nu_\mu+\bar{\nu}_\mu$ fluxes with the $\nu_\tau + \bar{\nu}_\tau$ 
fluxes being entirely suppressed \cite{ref46}, becomes marginal for 
neutrino energies above $10^5$ GeV \cite{ref19,ref21,ref22,ref31}, i.e.\
$E_\mu^{\rm min}=10^5$ GeV in Table 1, or, in other words, the ATM rate
comes entirely from $E_\nu<10^6$ GeV.

\section{Downward event rates}

For neutrino energies increasing beyond $10^5$ GeV the shadowing in Earth
rapidly increases (cf.~Figs.~2, 4 and Table 1) and eventually it becomes
beneficial to look for events induced by downward--going neutrinos. Since 
underground detectors are deployed at a depth of 2 to 4 km, the limited 
amount of matter above the detector does not induce any significant attenuation 
and regeneration of the initial cosmic neutrino fluxes \cite{ref19,ref21}, 
i.e., $F_{\stackrel{(-)}{\nu}\!\!\!_\tau}(E_\nu,X) 
\simeq F_{\stackrel{(-)}{\nu}\!\!\!_\tau}(E_\nu ,0) \equiv 
F_{\stackrel{(-)}{\nu}\!\!\!_\tau}^0(E_\nu)$
instead of (6).

Therefore the $\mu^{\pm}$ rates are calculated according to (12) with 
$F_{\stackrel{(-)}{\nu}\!\!\!_\tau}(E_\nu , X)$ replaced by 
$F_{\stackrel{(-)}{\nu}\!\!\!_\tau}^0(E_\nu)$.  Furthermore the lower 
limit of integration has to be raised at least to 
\mbox{$E_\mu^{\rm min}=10^5$ GeV} in order to suppress
the background due to atmospheric (ATM) and `prompt' $\nu_\mu +
\bar{\nu}_{\mu}$ fluxes \cite{ref46,ref48} (the atmospheric $\nu_\tau
+ \bar{\nu}_\tau$ flux is negligible and the prompt $\nu_\tau +
\bar{\nu}_\tau$ flux is about ten times smaller \cite{ref46} than the 
prompt $\nu_{\mu}+\bar{\nu}_\mu$ flux). For calculating `contained'
events, where the $\mu^{\pm}$ are produced by interactions within the
instrumented detector volume, we set $R(E_\mu , E_\mu^{\rm min})\equiv 1$
km$\,$we in (12) corresponding to an effective detector volume $V_{\rm eff}
= A_{\rm eff} \times 1$ km = 1 km$^3$ of water/ice in order to comply
\cite{ref2} with future underground detectors like IceCube and NEMO.
(A $\mu^{\pm}$ rate about ten times larger would be obtained if one uses
the analytic muon range (13), since the average value of $R$ is about
10 km we for $E_\nu>10^5$ GeV \cite{ref19,ref21}; the exploitation 
of this range enhancement of the effective volume is, however, illusory
since none of the future detectors will be deployed at a depth of 10 km.) 
These contained $\mu^- +\mu^+$ rates enhance by about 10\% (branching
fraction $B_{\mu}=0.18$ for $\tau\to\mu\nu_\mu\nu_\tau$) the $\mu^- +\mu^+$
event rates produced by the downward--going cosmic $\nu_\mu+\bar{\nu}_\mu$
flux \cite{ref19}, which are also shown (in parentheses) and needed in
Table 2 for the final total $\mu^- +\mu^+$ event rates.  Notice that
the downward muon event rates are larger by a factor of $2 -10$ than the
upward rates in Table 1 for $E_\nu >10^5$ GeV.  These results are
encouraging and allow to test some cosmic neutrino fluxes at higher
neutrino energies up to about $10^9$ GeV, in contrast to the 
upward--going events in Table 1 which are observable up to about $10^7$ GeV. 

In contrast to $\mu$--like events, hadronic `double bang' and `lollipop'
events are signatures unique to $\tau^{\pm}$ leptons produced by the
cosmic $\stackrel{(-)}{\nu}\!\!\!_\tau$ flux.  Furthermore, the atmospheric
flux background is negligible and the prompt $\nu_\tau +\bar{\nu}_\tau$
flux is about ten times smaller than the prompt $\nu_{\mu}+\bar{\nu}_\mu$
flux \cite{ref46}.  These specific hadronic event rates per unit solid
angle and second are calculated according to \cite{ref45}
\begin{equation}
N_h^{(\nu_\tau)} = B_{-\mu} A_{\rm eff} \int_{E_\tau^{\rm min}} dE_\nu\,
  P_h(E_\nu , E_\tau^{\rm min})\, F_{\nu_\tau}^0(E_\nu)
\end{equation}
with a reduction factor $B_{-\mu} = 1-B_\mu=0.82$ in order to exclude the
muonic mode of the $\tau$--decay, $A_{\rm eff} \simeq 1$ km$^2$ is the
effective area of the (underground) neutrino telescope, and $P_h$ is the
probability that the $\nu_\tau$ with energy $E_\nu$ produces a $h =$
`double bang' (db) event (i.e., two contained and separable hadronic showers)
or a $h =$ `lollipop' event (i.e., one hadronic shower arising from the 
semileptonic $\tau$--decay) with the $\tau$--energy greater than 
$E_\tau^{\rm min}$.  These two probabilities per incident tau--neutrino
are given by \cite{ref42}
\begin{eqnarray}
P_{\rm db}(E_\nu, E_\tau^{\rm min}) & = & 
  \rho N_A\int_0^{1-E_\tau^{\rm min}/E_\nu} dy\,\,
   \frac{d\sigma_{\nu_\tau N}^{\rm CC}(E_\nu, y)}{dy}\nonumber\\
& & \times\left[ (L_d-R_\tau^{\rm min}-R_\tau)e^{-R_\tau^{\rm min}/R_\tau}
   + R_\tau e^{-L_d/R_\tau}\right]\\
\nonumber\\
P_{\rm lollipop}(E_\nu , E_\tau^{\rm min})& = &
  \rho N_A (L_d-R_\tau^{\rm min}) \int_0^{1-E_\tau^{\rm min}/E_\nu}\!\!
    dy \,\,\frac{d\sigma_{\nu_\tau N}^{\rm CC}(E_\nu,y)}{dy}\, 
       e^{-R_\tau^{\rm min}/R_\tau}
\end{eqnarray}

\noindent 
with $\rho$ being the density of the detector medium ($\rho_{\rm ice}=
0.9$ g/cm$^3$), $L_d$ is the effective length scale of the detector
($L_d\simeq 1$ km) and the $\tau$--range $R_\tau =\lambda_\tau^{\rm dec}
(E_\tau)/\rho$, which must be contained within $L_d$, is given by
\begin{equation}
R_\tau(E_\nu ,y) = \frac{E_\tau}{m_\tau}\, c\tau_\tau =
    \frac{(1-y)E_\nu}{m_\tau}\, c \tau_\tau 
\end{equation}
with the constraint $R_\tau^{\rm min}\leq R_\tau\leq L_d$. The minimum
$\tau$--range $R_\tau^{\rm min}$ must be chosen so as to allow for shower
separation ($R_\tau^{\rm min}\simeq$ 100 m appears to be a reasonable 
effective value \cite{ref40,ref45} for IceCube where the horizontal
spacing of the photomultipliers \cite{ref2} is 125 m and their vertical
spacing is 16 m).  The lower limit of integration in (15) is taken to be
$E_\tau^{\rm min}=2\times 10^6$ GeV, since at this energy $R_\tau \simeq$
100 m which appears to allow for a clear separation of the two showers.
The upper limit of integration in (15) will be taken to be $10^{12}$ GeV
as usual.  However, for values 
$E_\nu$ \raisebox{-0.1cm}{$\stackrel{>}{\sim}$} $2\times 10^7$ GeV the 
$\tau$--range $R_\tau$ exceeds the assumed telescope size of $L_d\simeq 1$ 
km and thus double bang events become unobservable.  Although the 
probability for a lollipop event dominates \cite{ref42} over that for a
double bang for 
$E_\nu$ \raisebox{-0.1cm}{$\stackrel{>}{\sim}$} $5\times 10^6$ GeV, lollipop 
event rates become marginal for $E_\nu >10^8$ GeV as can be seen in 
Table 3.  The negligible background due to the atmospheric prompt
$\nu_\tau +\bar{\nu}_\tau$ flux~\cite{ref46} is also displayed in 
Table 3 for illustration. 
Despite being background--free and much larger than the $\nu_\tau +
\bar{\nu}_\tau$ induced upward and downward going $\mu$--like event rates
in Table 1 and 2 in the relevant neutrino energy range of $10^6$ to
$10^8$ GeV, these double bang and lollipop event rates are unique
signatures of cosmic $\nu_\tau + \bar{\nu}_\tau$ fluxes.

\section{Summary}
The importance and signatures of cosmic tau--(anti)neutrinos have been
analyzed for upward-- and downward--going $\mu^- +\mu^+$ and hadronic
shower event rates relevant for present and future underground water or
ice detectors.
The upward--going $\mu^- +\mu^+$ event rates initiated by cosmic 
$\nu_\mu +\bar{\nu}_\mu$ fluxes are enhanced by about 20 to 30\% by
taking into account cosmic $\nu_{\tau}+\bar{\nu}_\tau$ fluxes as well
as their associated $\tau^-+\tau^+$ fluxes.
In particular, 
the contributions arising from the $\tau^- +\tau^+$ flux are sizeable
and have been so far disregarded for calculating upward--going event
rates.  The different energy and nadir angle dependence of the
$\stackrel{(-)}{\nu}\!\!\!_\tau$ induced event rates may provide 
opportunities to identify these events among the multitude of 
$\stackrel{(-)}{\nu}\!\!\!_\mu$
induced events. Similarly the cosmic $\nu_\tau +\bar{\nu}_\tau$ fluxes
enhance the previously calculated $\nu_\mu +\bar{\nu}_\mu$ initiated
downward--going (contained) $\mu^- +\mu^+$ event rates by typically
about 10\% which allow to test some cosmic neutrino fluxes up to about
$10^9$ GeV -- two orders of magnitude higher than can be reached with
upward--going events.  In contrast to $\mu$--like events, downward--going
hadronic double bang and lollipop shower events are signatures unique
to $\tau^{\pm}$ leptons produced by cosmic $\nu_\tau +\bar{\nu}_\tau$
fluxes and, moreover, are background--free in the relevant energy 
region.  The rates are much larger than the $\nu_\tau +\bar{\nu}_\tau$
induced upward-- and downward--going $\mu$--like event rates in the 
relevant neutrino energy range of $10^6$ to $10^8$ GeV.  (Upward--going
double bang and lollipop event rates are small in the relevant energy
region $E_\nu>10^6$ GeV where the initial cosmic fluxes become strongly
attenuated and degraded in energy due to regeneration.)

For all our calculations we have used the nominal radiative GRV98 parton
distributions with their unique QCD--dynamical small--$x$ predictions.
It should be noticed that the relevant CC and NC cross sections obtained
from the  `variable flavor' CTEQ3--DIS parton densities with their
assumed fixed--power extrapolation to $x<10^5$ accidentally coincide
practically with the ones derived from the dynamical ultrasmall--$x$
predictions of the radiative parton model.  In contrast to the 
upward--going event rates, the downward--going rates for ultrahigh
neutrino energies depend strongly on the specific choice of parton
distributions and their behavior in the ultrasmall Bjorken--$x$ region.

In order to estimate upward--going event rates one has to deal with
coupled transport equations for $\stackrel{(-)}{\nu}\!\!\!_\tau$ fluxes
and their associated $\tau^{\pm}$ fluxes.  Since we do not fully confirm
the quantitative results for the $\nu_\tau +\bar{\nu}_\tau$ flux
obtained in the literature so far, the solutions of the coupled transport
equations are recapitulated for completeness in the Appendix, together
with the approximations relevant for our calculations.  The typical
enhancement (`bump') of the upward--going attenuated and regenerated 
$\stackrel{(-)}
{\nu}\!\!\!_\tau$ flux around $10^4 - 10^5$ GeV amounts to about 40\%
with respect to the initial cosmic flux, which is prominent for flatter
initial cosmic fluxes 
$F_{\stackrel{(-)}{\nu}\!\!\!_\tau}^0 \sim E_\nu^{-1}$. This is in contrast to
an enhancement of about a factor of 2 found previously.  The related
upward--going event rates are therefore, on the average, about 50\%
smaller than previously estimated.  On the other hand, for steeper
initial cosmic fluxes $F_{\stackrel{(-)}{\nu}\!\!\!_\tau}^0 \sim E_{\nu}^{-n}$,
\mbox{$n$ \raisebox{-0.1cm}{$\stackrel{>}{\sim}$} 2}, the differences are 
always less than 10\%.

This work has been supported in part by the `Bundesministerium f\"ur
Bildung und Forschung', Berlin/Bonn. 
 
\newpage

\appendix
\section*{Appendix}
\renewcommand{\theequation}{A.\arabic{equation}}
\setcounter{equation}{0}
In order to solve the coupled integro--differential equations (1) and (2),
it is convenient to solve first (2) and to rewrite it as
\begin{equation}
\left[ \frac{\partial}{\partial X}-\gamma(E)\frac{\partial}{\partial E}
 + A(E)\right] F_{\tau}(E,X) = G_\nu(E,X)
\end{equation}
with 
$A(E)=1/\hat{\lambda}(E)-\partial\gamma(E)/\partial E$ and
$G_\nu(E,X) = \lambda_{\nu_\tau}^{-1}(E)\int_0^1\frac{dy}{1-y}\,
 K_{\nu_\tau}^{\rm CC}(E,y)\, F_{\nu_\tau}(E_y,X)$.
The homogeneous equation, i.e.\ for $G_\nu\equiv 0$, being similar to
the well known renormalization group equation of asymptotic Green's
functions (see, e.g.\ \cite{ref65}), can be solved by the usual ansatz
\begin{equation} 
F_\tau(E,X) = f(E,X)\exp 
     \left[ \int_0^E \frac{A(E')}{\gamma(E')}\, dE'\right]
\end{equation}
in order to remove the nonderivative $A$--term in (A.1), which leads
to
\begin{equation}
\left[\frac{\partial}{\partial X}-\gamma(E)
                \frac{\partial}{\partial E}\right]f(E,X) = 0\, .
\end{equation}
This equation can be solved by introducing, as usual, an effective
  `running' energy $\bar{E}(X,E)$ defined by
\begin{equation}
\frac{d}{dX}\bar{E}(X,E) = \gamma(\bar{E})\,\, ,\quad\quad
   \bar{E}(0,E) = E\, ,
\end{equation}
in order to satisfy the same differential equation (A.3) for $f(E,X)$,
\begin{equation}
\left[\frac{\partial}{\partial X}-\gamma(E) 
   \frac{\partial}{\partial E}\right]\, \bar{E}(X,E)=0\, .
\end{equation}
Thus if $f$ depends on $X$ and $E$ through the combination 
$\bar{E}(X,E)$, i.e.\ $f(E,X)=f\left( \bar{E}(X,E),0\right)$, it
will satisfy (A.3) and the homogeneous solution (A.2) becomes
\begin{equation}
F_\tau(E,X) = f\left( \bar{E}(X,E),0 \right)\exp 
  \left[\int_0^E \frac{A(E')}{\gamma(E')}\,\,  dE'\right]
\end{equation}
or, using (A.4),
\begin{equation}
F_\tau(E,X) = F_\tau\left( \bar{E}(X,E),0\right)
   \exp\left[-\int_0^X A\left(\bar{E}(X',E)\right)\, dX'\right]\, .
\end{equation}
The solution of the full inhomogeneous ($G_\nu \neq 0$) equation
(A.1) is then commonly written as \cite{ref61}
\begin{equation}
F_\tau(E,X) =\int_0^X dX'G_\nu\left( \bar{E}(X-X',E),\, X'\right)
 \exp \left[-\int_{X'}^X A\left( \bar{E}(X-X'',E)\right)\, dX''\right]\, .
\end{equation}

Next, Eq.~(1) can be solved by the ansatz $\left( {\rm cf.}(6)\right)$
\begin{equation}
F_{\nu_\tau}(E,X)= F_{\nu_\tau}^0(E) \exp 
      \left[ -\frac{X}{\Lambda_\nu (E,X)}\right]
\end{equation}
with an `effective interaction (absorption) length' \cite{ref30}
\begin{equation}
\Lambda_{\nu}(E,X) = \frac{\lambda_{\nu_\tau}(E)}{1-Z(E,X)}
\end{equation}
which, when inserted into (1), yields
\begin{eqnarray}
XZ(E,X) & = & \int_0^X dX'\!\!\int_0^1 dy \, K_{\nu_\tau}^{\rm NC}(E,y)
 \eta _{\nu_\tau}(E,y)\, e^{-X'D_\nu(E,E_y,X')}\nonumber\\
& & + \lambda_{\nu_\tau}(E)\int_0^X \!\! dX'\int_0^1 \!\! dy\, K_\tau(E,y)\, 
   F_\tau(E_y,X')\frac{\eta_{\nu_\tau}(E,y)}{F_{\nu_\tau}^0(E_y)}\,\,
      e^{X'/\Lambda_\nu(E,X')}\nonumber\\
&& 
\end{eqnarray}
with $D_\nu(E,E_y,X') = \Lambda_\nu^{-1}(E_y,X')-\Lambda_\nu^{-1}(E,X')$.
Note that this $Z$--factor also appears in $F_\tau$ where it enters via
$G_\nu$ in (A.8) which is proportional to $F_{\nu_\tau}$ 
$\left( {\rm c.f.}\, ({\rm A}.1)\right)$. It is convenient to solve for 
$Z(E,X)$ iteratively \cite{ref30}, starting with $Z^{(0)}(E,X)=0$ on 
the r.h.s. of (A.9) and (A.11), which yields the sufficiently accurate 
first iterative solution $Z^{(1)}$.  (This iteration procedure converges very
quickly: the difference between the second iteration $Z^{(2)}$ and
$Z^{(1)}$ is negligible \cite{ref44} (and deviates at most by 4\% from
$Z^{(1)}$ in the case of an upward--going $\nu_{\mu}+\bar{\nu}_\mu$
flux \cite{ref30,ref66}) for the initial cosmic neutrino fluxes under
consideration in Fig.~1.)  From (A.11) we get
\begin{eqnarray}
XZ^{(1)}(E,X) & =& \int_0^1 dy\, K_{\nu_\tau}^{\rm NC}(E,y)\,
    \eta_{\nu_\tau}(E,y)\, \frac{1-e^{-XD_\nu(E,E_y)}}{D_\nu(E,E_y)}
\nonumber\\
& & +\lambda_{\nu_\tau}(E)\int_0^1 dy\, K_\tau(E,y)\, 
    \frac{\eta_{\nu_\tau}(E,y)}{F_{\nu_\tau}^0(E_y)} \int_0^X dX'
      F_\tau^{(0)}(E_y,X')\, e^{X'/\lambda_{\nu_\tau}(E)}
\nonumber\\
& \equiv & X(Z_\nu^{(1)}+Z_\tau^{(1)})
\end{eqnarray}
where we have used $\Lambda_\nu^{(0)}(E,X')=\lambda_{\nu_\tau}(E)$ and
$D_\nu^{(0)}(E,E_y,X') = D_\nu(E,E_y)$ with $D_\nu(E,E_y)$ given in (9),
and $Z_\nu^{(1)}$ is the expression given in (8).  Furthermore the 
required $F_\tau^{(0)}(E_y,X')$ in $Z_\tau^{(1)}$ in (A.12) follows
from (A.8) with
\begin{equation} 
G_\nu^{(0)}\left( \bar{E}(X-X',E),X'\right) = 
  \frac{1}{\lambda_{\nu_\tau}(\bar{E})} \int_0^1\frac{dy}{1-y}\,
    K_{\nu_\tau}^{\rm CC}(\bar{E},y)\, F_{\nu_\tau}^0(\bar{E}_y)\, 
      e^{-X'/\lambda_{\nu_\tau}(\bar{E}_y)}
\end{equation}
with $\bar{E}_y=\bar{E}(X-X',E_y)$. In contrast to $Z_\nu^{(1)}$ in
(A.12), the $X$--integrals in $F_\tau^{(0)}$ and in $Z_\tau^{(1)}$ in 
(A.12)
cannot be further simplified analytically.  Having obtained $Z^{(1)}=
Z_\nu^{(1)}+Z_\tau^{(1)}$, the first iteration $\nu_\tau$--flux 
$F_{\nu_\tau}^{(1)}(E,X)$ is given by (A.9) with
$\Lambda_\nu^{(1)}=\lambda_{\nu_\tau}/(1-Z^{(1)})$ which generates the
$\tau$--flux $F_\tau^{(1)}(E,X)$ via (A.8) where 
\begin{equation}
G_\nu^{(1)}\left( \bar{E}(X-X',E),X'\right) 
  = \frac{1}{\lambda_{\nu_\tau}(\bar{E})}\int_0^1 \frac{dy}{1-y}
   K_{\nu_\tau}^{\rm CC}(\bar{E},y) F_{\nu_\tau}^{(1)}(\bar{E}_y,X')\, .
\end{equation}

If, however, the $\tau^{\pm}$ energy--loss can be neglected, 
$\gamma(E_\tau)\simeq 0$ according to (5) (or, alternatively, if the 
$\tau^{\pm}$ energy--loss is treated separately \cite{ref57,ref59} in
which case the term proportional to $\gamma(E)$ in the transport
equation (2) is absent from the very beginning), the complicated
$X$--integrals in $F_\tau^{(0)}$ and $Z_\tau^{(1)}$ can be performed 
analytically.  Since in this approximation $\bar{E}=E$ and in (A.1)
$A(E)\simeq 1/\lambda_\tau^{\rm dec}(E)$ according to (5), one obtains
from (A.8) and (A.13) the lowest order $\tau$--flux
\begin{equation}
F_\tau^{(0)}(E,X) = \frac{F_{\nu_\tau}^0(E)}{\lambda_{\nu_\tau}(E)}
\,\, e^{-\frac{X}{\lambda_\tau^{\rm dec}(E)}} \int_0^1 dy'\, 
  K_{\nu_\tau}^{\rm CC}(E,y')\eta_{\nu_\tau}(E,y')\, 
   \frac{1-e^{-XD_{\nu\tau}(E,E_{y'})}}{D_{\nu\tau}(E,E_{y'})}
\end{equation}
with $D_{\nu\tau}(E,E_{y'})$ defined in (9).  Inserting (A.15) into 
(A.12) and using 
$K_\tau(E,y)\simeq K_\tau^{\rm dec}(E,y)/\lambda_\tau^{\rm dec}(E)$,
according to the approximation (5), results in the expression for 
$Z_\tau^{(1)}$
given in (8).  Together with $Z_\nu^{(1)}$ in (A.12), this finally
determines $Z^{(1)}=Z_\nu^{(1)}+Z_\tau^{(1)}$ and thus the first
iteration $\nu_\tau$--flux $F_{\nu_\tau}^{(1)}$ via (A.9) which has
been denoted for simplicity by $F_{\nu_\tau}$ in (6).  This first 
order $\nu_\tau$--flux $F_{\nu_\tau}^{(1)}$ now generates the $\tau$
--flux $F_\tau^{(1)}$ via (A.8),
\begin{eqnarray}
F_\tau^{(1)}(E,X) & = & e^{-\frac{X}{\lambda_\tau^{\rm dec}(E)}}
   \int_0^X dX'\, G_\nu^{(1)}(E,X')\, 
    e^{\frac{X'}{\lambda_\tau^{\rm dec}(E)}}\nonumber\\
& = & 
\lambda_{\nu_\tau}^{-1}(E)\, e^{-\frac{X}{\lambda_\tau^{\rm dec}(E)}}
   \!\!\!\int_0^X \!\!\! dX'\!\!\!
    \int_0^1\!\!\!\frac{dy}{1-y} K_{\nu_\tau}^{\rm CC}(E,y)
 F_{\nu_\tau}^{(1)}(E_y,X')\, e^{\frac{X'}{\lambda_\tau^{\rm dec}(E)}}
\end{eqnarray}
which, using (A.9) for $F_{\nu_\tau}^{(1)}$, is the expression given in (7) where, for 
simplicity, this first iteration flux has been denoted by
$F_\tau(E,X)$.

\raggedbottom

\newpage

\pagestyle{empty}

\setlength{\oddsidemargin}{-1.0cm}
\begin{table}
\renewcommand{\arraystretch}{1.2}

\parbox{18cm}{\footnotesize{Table 1: Total nadir--angle--integrated upward--going $\mu^- +\mu^+$
       event rates per year from $(\nu_\tau+\bar{\nu}_\tau)N$ and 
       $(\nu_\mu+\bar{\nu}_\mu)N$ interactions in rock, with the latter
       being given in parentheses which are taken from Table 1 of
       \cite{ref19}, for various muon energy thresholds 
       $E_\mu^{\rm min}$ and the appropriate cosmic neutrino fluxes
       in Fig.~1.  The $\nu_\tau+\bar{\nu}_\tau$ initiated rates are
       calculated according to Eqs.~(12) and (14), multiplied by
       $2\pi$, i.e.\ $2\pi(N_{\mu^- +\mu^+}^{(\nu_\tau+\bar{\nu}_\tau)}
       + N_{\mu^- +\mu^+}^{(\tau^-+\tau^+)})$ as given by the solid
       curves in Fig.~6, and added to the total $\nu_\mu+\bar{\nu}_\mu$
       initiated rates in parentheses in order to obtain the final
       total rates. A `bar' signals that the rates fall below 0.01.}}

\vspace{0.5cm}

\begin{tabular}{|l|l||l|l|l|l|l|}
\hline
\raisebox{-1.5ex}[-1.5ex]{Flux} &\raisebox{-1.5ex}[-1.5ex]{Detector}& 
\multicolumn{5}{c|}{Muon-energy threshold E$_{\mu}^{\rm{min}}$/GeV} \\
      &        & $10^3$& $10^4$& $10^5$ & $10^6$& $10^7$\\ \hline\hline
      & ANTARES&   525 (411)&308 (248)&105.54 (89.3)&14.95 (13.0)&0.56 (0.53)\\ 
AGN-SS& AMANDA-II& 910 (699)&512 (408)&162 (137)&21.67 (19.3)&0.84 (0.79)\\ 
      &IceCube&    3534 (2687)&1945 (1547)&609 (514)&81.52 (72.6)&3.18 (3.00)\\ 
\hline
       & ANTARES&  16.45 (13.7)&6.18 (5.00)&2.47 (1.98)&1.05 (0.90)&0.34 (0.32)\\
AGN-M95&AMANDA-II& 34.59 (29.1)&10.57 (8.62)&3.72 (2.98)&1.56 (1.34)&0.48 (0.46)\\      
       & IceCube&  169 (143)&41.22 (33.7)&13.98 (11.2)&5.87 (5.04)&1.82 (1.74)\\
\hline
      & ANTARES&  0.74 (0.60)&0.38 (0.32)&0.09 (0.08)&0.01 (0.01)&-- \\ 
GRB-WB&AMANDA-II& 1.38 (1.10)&0.68 (0.56)&0.15 (0.13)&0.02 (0.02)&-- \\  
      & IceCube&  5.54 (4.35)&2.59 (2.13)&0.57 (0.49)&0.07 (0.06)&-- \\
\hline
      & ANTARES&  0.82 (0.62)&0.58 (0.45)&0.32 (0.26)&0.14 (0.12)&0.05 (0.05)\\ 
TD-SLBY&AMANDA-II&1.30 (0.97)&0.88 (0.68)&0.48 (0.39)&0.21 (0.18)&0.07 (0.07)\\  
      & IceCube&  4.96 (3.70)&3.34 (2.57)&1.81 (1.47)&0.77 (0.68)&0.26 (0.25)\\
\hline
      & ANTARES&  0.01 (0.01)&0.01 (0.01)&-- &-- &-- \\
TD-SLSC&AMANDA-II&0.01 (0.01)&0.01 (0.01)&0.01 (0.01)&-- &-- \\  
      & IceCube&  0.05 (0.04)&0.04 (0.03)&0.02 (0.02)&0.01 (0.01)&0.01 (0.01)\\
\hline
      & ANTARES&  0.01 (0.01)&0.01 (0.01)&0.01 (0.01)&0.01 (0.01)&-- \\ 
Z-burst&AMANDA-II& 0.01 (0.01)&0.01 (0.01)&0.01 (0.01)&0.01 (0.01)&0.01 (0.01)\\  
      & IceCube&   0.05 (0.05)&0.04 (0.04)&0.03 (0.03)&0.03 (0.03)&0.02 (0.02)\\ 
\hline
\end{tabular}
\end{table}

\clearpage
\setlength{\oddsidemargin}{0.0cm}

%
\begin{table}
\centering
\renewcommand{\arraystretch}{1.2}
\parbox{17cm}{\footnotesize{Table 2: Total (contained) downward $\mu^- +\mu^+$ event rates per 
       year for an effective detector volume $V_{\rm eff}\equiv 
       A_{\rm eff}\times 1$ km = 1 km$^3$ of water/ice. 
       The $\nu_\tau +\bar{\nu}_\tau$ initiated rates are calculated
       according to Eq.~(12), multiplied by $2\pi$, as explained in
       the text, and added to the total $\nu_\mu +\bar{\nu}_\mu$
       initiated rates in parentheses (which are taken from Table 2
       of \cite{ref19}) in order to obtain the final total rates.
       The background event rates are due to the dominant conventional
       atmospheric (ATM) \cite{ref48} and   `prompt' \cite{ref46}
       $\nu_\mu +\bar{\nu}_\mu$ fluxes.  (Notice that, since the 
       atmospheric and prompt $\nu_\tau+\bar{\nu}_\tau$ fluxes are
       negligible \cite{ref46}, the final total rates and the ones in
       parentheses coincide here.) A  `bar' signals that the rates 
       fall below 0.01}.}
\vspace{0.5cm}

\begin{tabular}{|l|l|l|l|l|l|l|}
\hline
      & \multicolumn{6}{c|}{$E_\mu^{\rm{min}}$ [GeV]}  \\
\raisebox{1.5ex}{flux} & $10^5$ & $10^6$ & $10^7$ & $10^8$ & $10^9$ & $10^{10}$ \\
\hline\hline
ATM & 1.15 (1.15) &0.01 (0.01) &--&--&--&--           \\
Prompt  &0.58 (0.58) &0.03 (0.03) & -- & -- &--&--   \\
AGN-SS  & 560 (510)  & 220 (207) &31.86 (30.9) &0.34 (0.34)   &-- &  --     \\
AGN-M95 &13.47 (11.8)&10.27 (8.95)&7.91 (7.09)&4.03 (3.74)   &0.92 (0.88) 
&0.05 (0.05)    \\
GRB-WB  &0.66 (0.61) &0.17 (0.16)   &0.02 (0.02)   &-- &--&-- \\
TD-SLBY &1.71 (1.47)&1.49 (1.30)&1.13 (1.00)&0.70 (0.63)&0.33 (0.30) 
&0.11 (0.10)  \\
TD-SLSC &0.03 (0.03) &0.03 (0.03)&0.03 (0.03)&0.02 (0.02)  &0.02 (0.02)  & 
0.01 (0.01)    \\
Z-burst &0.09 (0.08)&0.09 (0.08)&0.09 (0.08) &0.09 (0.08)   &0.09 (0.08)  
&0.07 (0.06)    \\
\hline
\end{tabular}
\end{table}
\vspace{1.5cm}
\begin{table}
\centering
\renewcommand{\arraystretch}{1.2}
\parbox{14cm}{\footnotesize{Table 3: Total downward--going double bang and lollipop event rates
       per year initiated by the cosmic $\nu_\tau+\bar{\nu}_\tau$
       fluxes in Fig.~1 and calculated according to Eq.~(15), multiplied
       by $2\pi$, as explained in the text for an IceCube--like
       km$^3$--sized detector.  The negligible background is due to
       the atmospheric prompt $\nu_\tau+\bar{\nu}_\tau$ flux \
       \cite{ref46}.  A `bar' signals that the rates fall below 0.01}.}

\vspace{0.5cm}

\begin{tabular}{|l|l|l|l|l|l|l|}
\hline
       &\multicolumn{3}{c|}{$N_{\mathrm{double~bang}}$}
       &\multicolumn{3}{c|}{$N_{\rm{lollipop}}$}  \\
\cdashline{2-7}
       & \multicolumn{3}{c|}{$E_\tau^{\rm{min}}$ [GeV]}
       &\multicolumn{3}{c|}{$E_\tau^{\rm{min}}$ [GeV]}  \\
 flux &$2 \times 10^6$ & $10^7$ & $10^8$ &$2 \times 10^6$ & $10^7$ & $10^8$ \\
\hline\hline
Prompt &$6\times 10^{-5}$ &$6\times 10^{-6}$  &$6\times 10^{-9}$&
$9\times 10^{-5}$&$2\times 10^{-5}$&$6\times 10^{-8}$           \\
AGN-SS  & 28.15 & 4.73  &0.01  &43.92&12.89 &  0.13     \\
AGN-M95 &1.07 &0.59 &0.07 &4.84 &4.15&2.09     \\
GRB-WB  &0.02  &-- &-- &0.03 &0.01&-- \\
TD-SLBY &0.13 &0.07 &0.01 &0.63 &0.60  &0.37   \\
TD-SLSC &--  &-- &-- &0.02   &0.02   & 0.01     \\
Z-burst &-- &-- &--  &0.05 &0.05   &0.05     \\
\hline
\end{tabular}
\end{table}


\clearpage

\begin{figure}
\centering
\includegraphics[width=0.7\textwidth,angle=270]{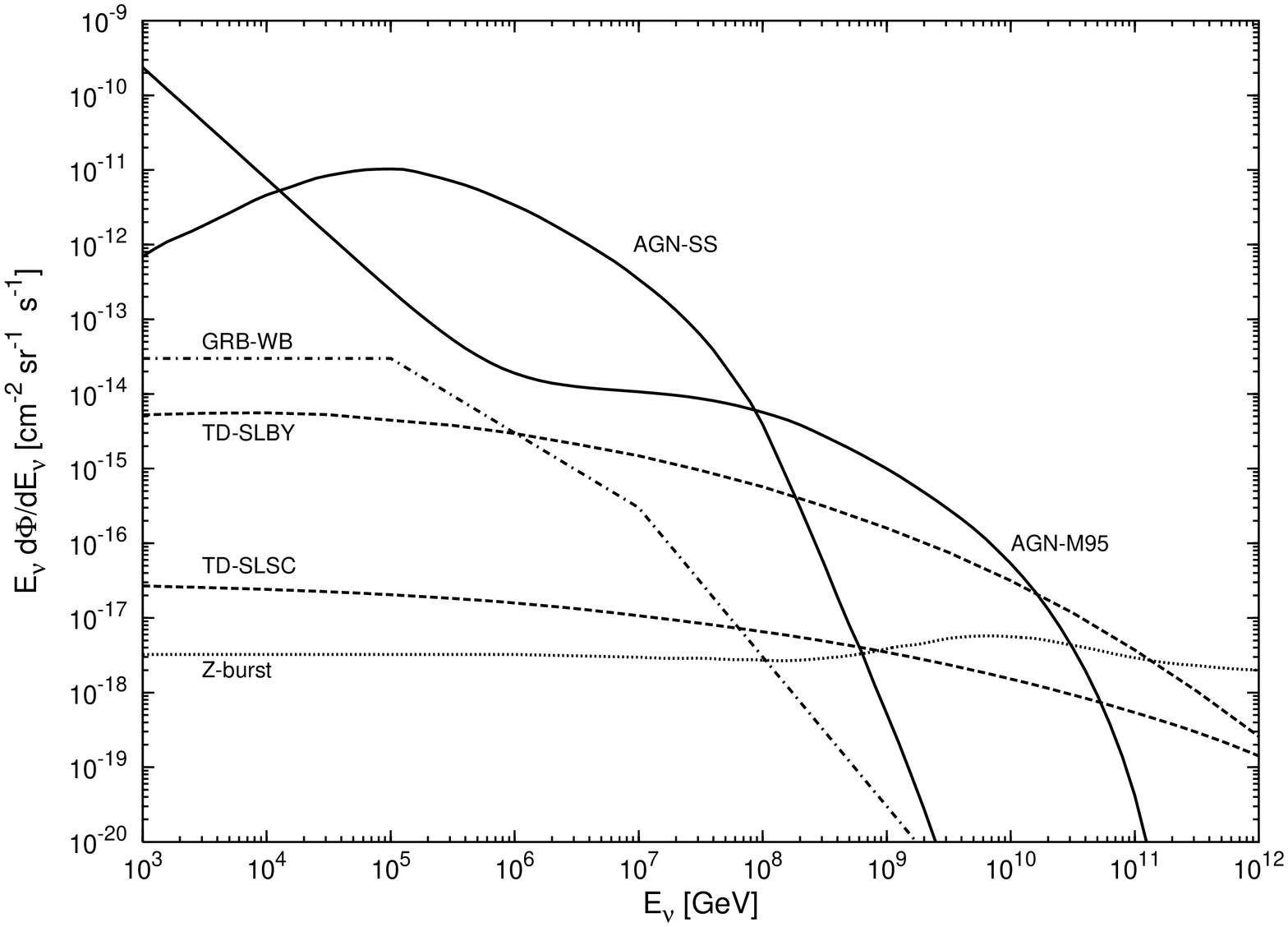}
%
\parbox{15cm}{\medskip
      \footnotesize{Figure 1: Representative differential fluxes of muon neutrinos 
      ($\nu_\mu+\bar{\nu}_\mu$) from active galactic nuclei (AGN--SS
      \cite{ref7} and AGN--M95 \cite{ref6}), gamma ray bursts
      (GRB--WB \cite{ref9}), topological defects (TD--SLSC \cite{ref12}
      and TD--SLBY \cite{ref13}) and $Z$--bursts \cite{ref17}. Due
      to naive channel counting in pion production and decay at the 
      production site ($\nu_e:\nu_\mu :\nu_\tau = 1:2:0$) and maximal
      mixing, $\nu_e:\nu_\mu :\nu_\tau = 1:1:1$, these fluxes are 
      divided equally between $e$--, $\mu$-- and $\tau$--neutrinos 
      when they reach the Earth's surface (i.e.\ will be divided by
      a factor of 2).  Notice that the AGN--SS flux is in conflict
      with a recent upper bound from the AMANDA--B10 detector 
      \cite{ref20} for $10^6$ GeV $\leq E_\nu <10^8$ GeV.}}
\end{figure}

\newpage

\begin{figure}
\includegraphics[width=\textwidth]{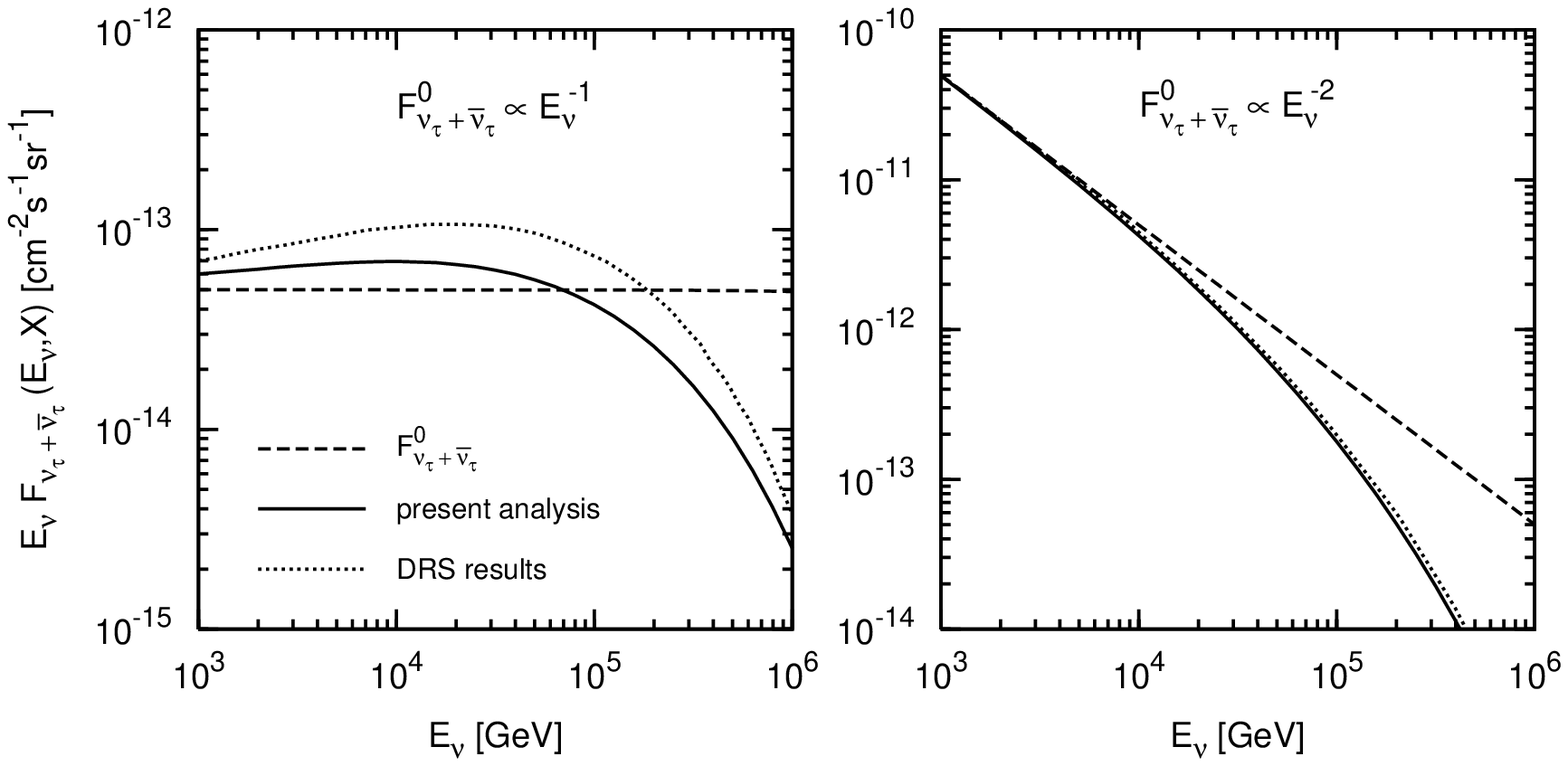}
\parbox{16cm}{\medskip
      \footnotesize{Figure 2: Attenuated and regenerated $\nu_\tau+\bar{\nu}_\tau$ fluxes
      calculated according to (6) and (8) for a nadir angle $\theta
      = 0^{\rm o}$ using the initial fluxes in (10) and (11).  For
      comparison the DRS results \cite{ref33} are shown as well.}} 
\end{figure}

\begin{figure}
\includegraphics[width=\textwidth]{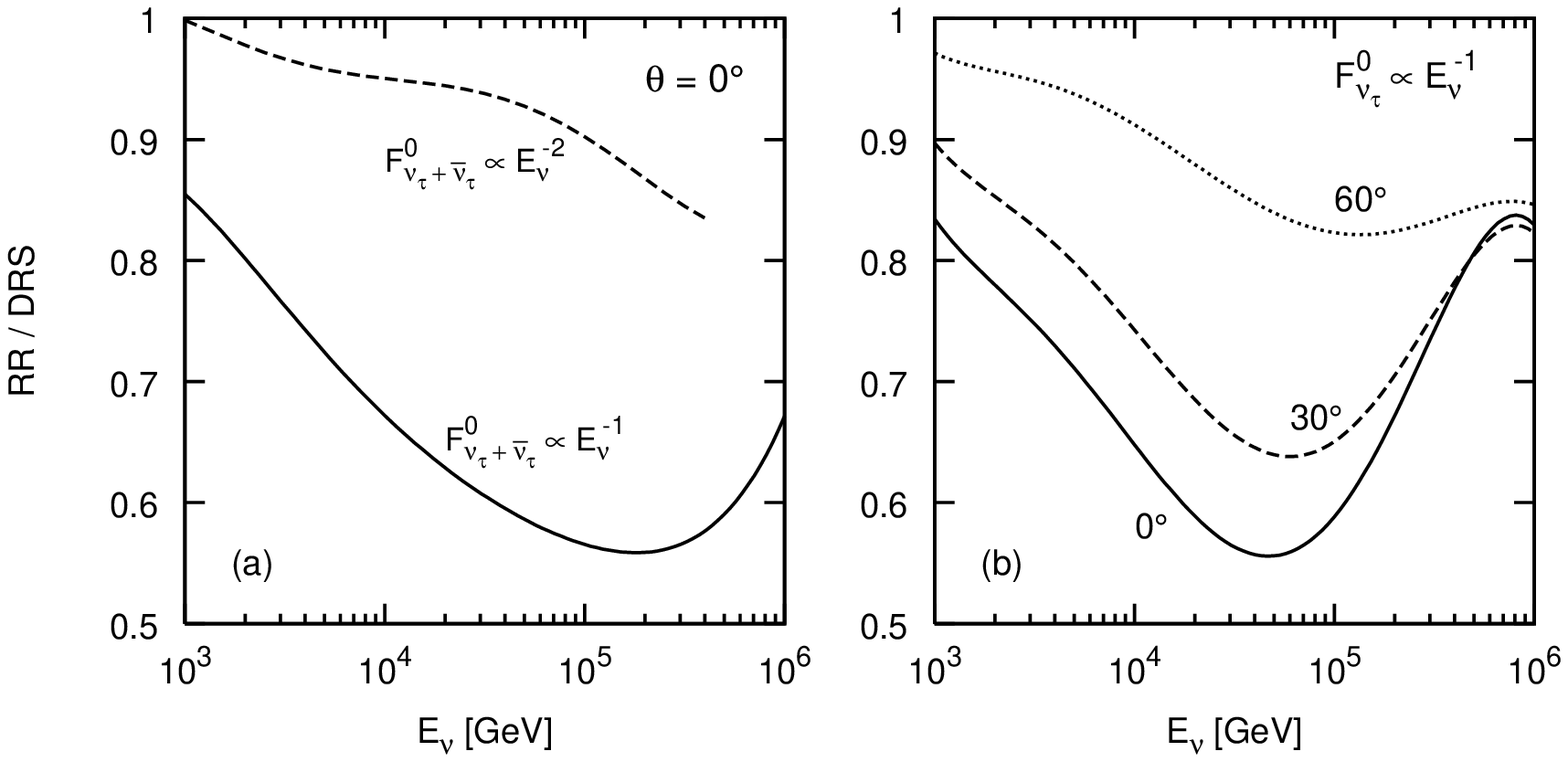}
\parbox{16cm}{\medskip
      \footnotesize{Figure 3: Ratios of our (RR) results and (a) the ones of DRS 
      \cite{ref33}, which correspond to the results in Fig.~2, and
      (b) the DRS ones of \cite{ref62}, which are the same as in 
      \cite{ref32}, for some representative values of the nadir
      angle $\theta$. The ratios for $F_{\nu_\tau +\bar{\nu}_\tau}^0
      \sim E_\nu^{-2}$ at $\theta >0^{\rm o}$ are always smaller 
       than the one in (a) for $\theta =0^{\rm o}$.}}
\end{figure} 

\begin{figure}
\centering
\vspace{-2.0cm}
\includegraphics[width=\textwidth]{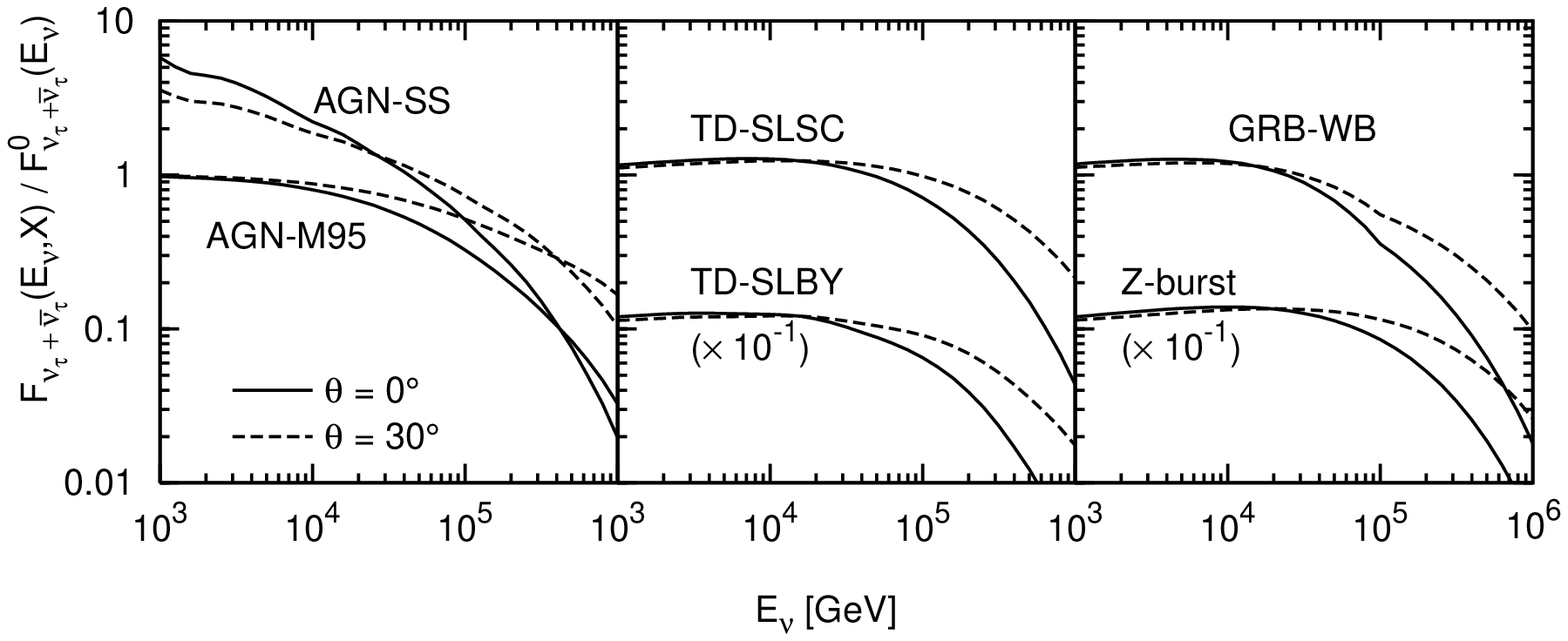}
\parbox{16cm}{\medskip
      \footnotesize{Figure 4: Ratios of the attenuated and regenerated 
      $\nu_\tau+\bar{\nu}_\tau$ fluxes at $\theta = 0^{\rm o}$, 
      $30^{\rm o}$ and the initial cosmic fluxes 
      $F_{\nu_\tau +\bar{\nu}_\tau}^0$ in Fig.~1.  The results for
      the TD--SLBY and $Z$--burst fluxes are multiplied by $10^{-1}$
      as indicated.}}
\end{figure}

\begin{figure}
\vspace{-2.0cm}
\includegraphics[width=\textwidth]{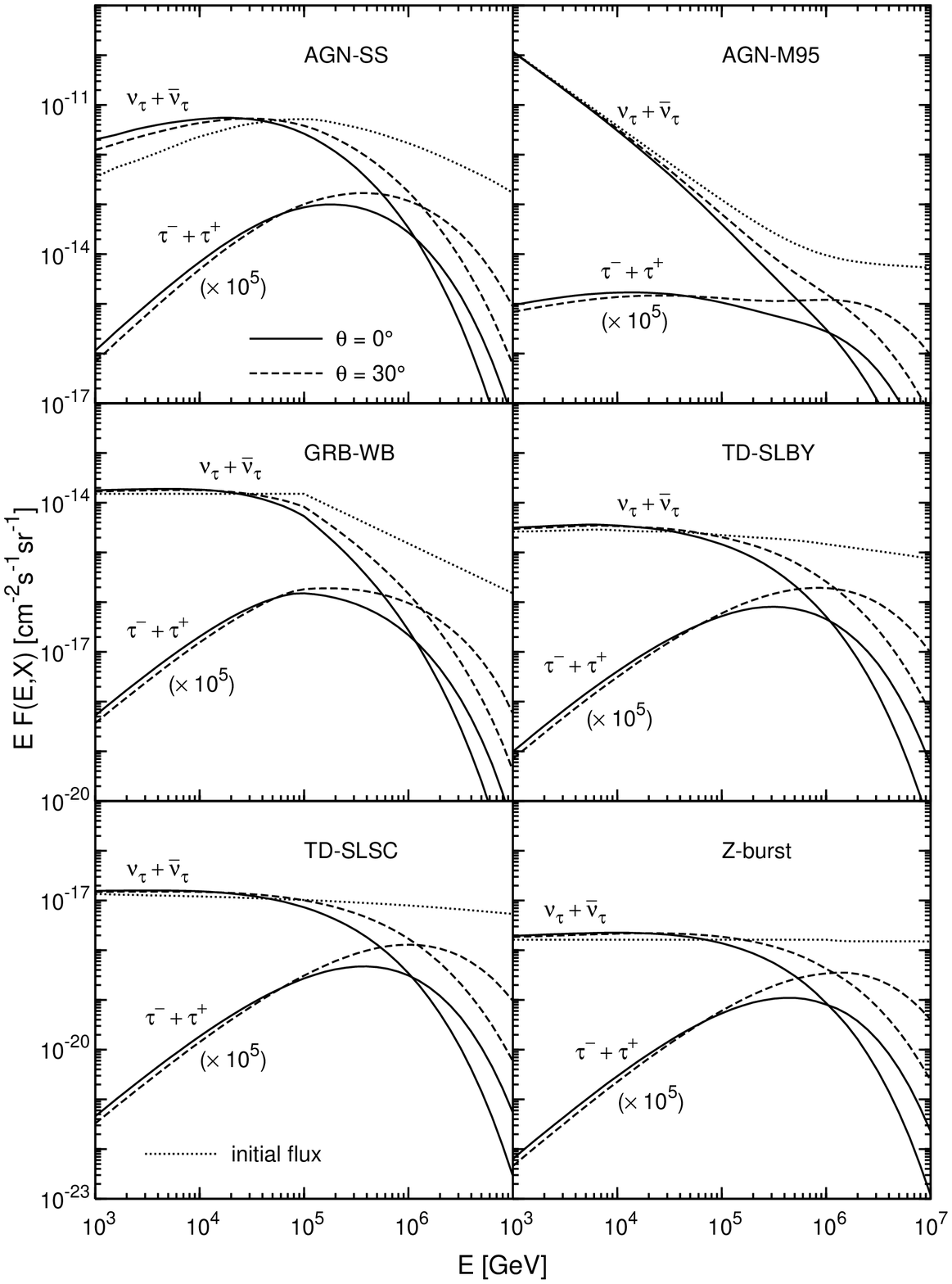}
\parbox{16cm}{\medskip
      \footnotesize{Figure 5: Attenuated and regenerated $\nu_\tau +\bar{\nu}_\tau$ and
      $\tau^- +\tau^+$ fluxes calculated according to (6) and (7),
      respectively, for nadir angles $\theta =0^{\rm o}$ and 
      30$^{\rm o}$ using the initial fluxes 
      $F_{\nu_\tau +\bar{\nu}_\tau}^0 =\frac{1}{2}\, d\Phi/dE_\nu$ 
      in Fig.~1 which are shown by the dotted curves.  All results
      for the $\tau^- +\tau^+$ fluxes are multiplied by $10^5$ as
      indicated.}}
\end{figure}

\begin{figure}
\includegraphics[width=\textwidth]{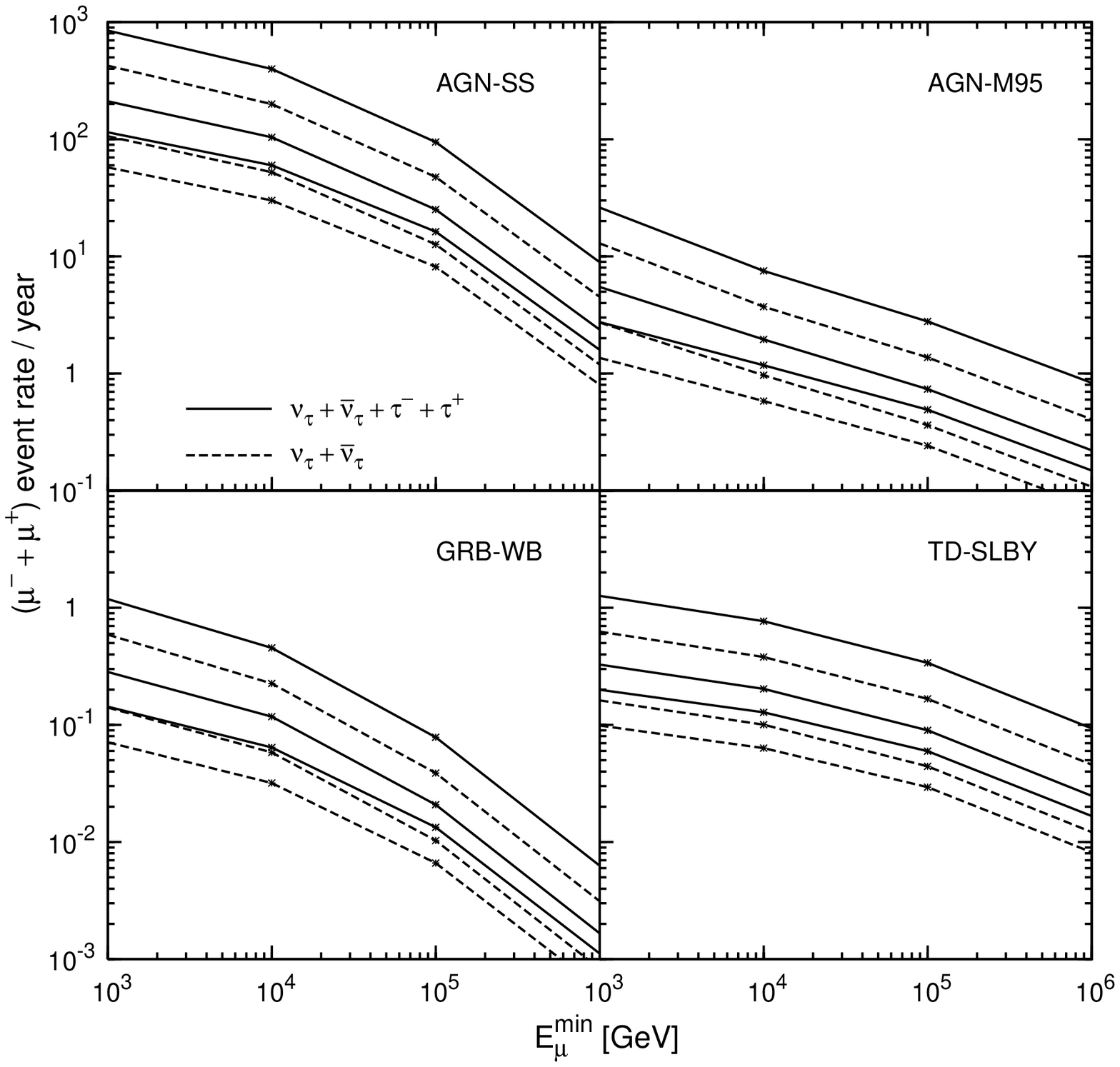}
\parbox{16cm}{\medskip
      \footnotesize{Figure 6: Total nadir--angle--integrated upward--going $\mu^- +\mu^+$
      event rates per year initiated by the initial cosmic 
      $\nu_\tau +\bar{\nu}_\tau$ fluxes in Fig.~1 as a function of 
      $E_\mu^{\rm min}$.  The rates arising just from the 
      $\nu_\tau +\bar{\nu}_\tau$ fluxes (dashed curves) are 
      calculated according to (12), multiplied by $2\pi$; adding
      the rates arising from the $\tau^- +\tau^+$ flux according to
      (14), multiplied by $2\pi$, one obtains the total rates shown
      by the solid curves.  The largest rates shown by the solid and
      dashed curves refer to the IceCube and, in decreasing order,
      to the \mbox{AMANDA--II} and ANTARES underground detectors, 
      respectively.}}
\end{figure}
\end{document}